\documentclass[aps,pre,floats,twocolumn,superscriptaddress]{revtex4}

\usepackage{graphicx,epsfig}
\usepackage{times}
\usepackage{graphics,dcolumn,bm,fleqn,epic,eepic,float}
\usepackage{amssymb,amsmath,multirow,rotate,color}
\begin{document}

\title{Spreading of sexually transmitted diseases in heterosexual populations}
\author{Jes\'us G\'omez-Garde\~nes}

\affiliation{Scuola Superiore di Catania, Via S. Paolo 73, 
95123 Catania, Italy}

\affiliation{Institute for Biocomputation and Physics of Complex
Systems (BIFI), University of Zaragoza, Zaragoza 50009, Spain}

\author{ Vito Latora}

\affiliation{Dipartimento di Fisica e Astronomia, Universit\`a di Catania, 
and INFN, Via S. Sofia 64, 95123 Catania, Italy}

\author{Yamir Moreno}

\affiliation{Institute for Biocomputation and Physics of Complex
Systems (BIFI), University of Zaragoza, Zaragoza 50009, Spain}

\author{Elio V. Profumo}

\affiliation{Scuola Superiore di Catania, Via S. Paolo 73, 
95123 Catania, Italy}

\date{\today}

\begin{abstract}
The spread of sexually transmitted diseases ({\em e.g. Chlamydia, Syphilis,
Gonorrhea, HIV}) across populations is a major concern for
scientists and health agencies.  In this context, both data collection
on sexual contact networks and the modeling of disease spreading, are
intensively contributing to the search for effective immunization
policies. Here, the spreading of sexually transmitted diseases on
bipartite scale-free graphs, representing heterosexual contact
networks, is considered.  We analytically derive the expression for
the epidemic threshold and its dependence with the system size in
finite populations.  We show that the epidemic outbreak in bipartite
populations, with number of sexual partners distributed as in
empirical observations from national sex surveys, takes place for
larger spreading rates than for the case in which the bipartite nature
of the network is not taken into account.  Numerical simulations
confirm the validity of the theoretical results.  Our findings 
indicate that the restriction to crossed infections between the two
classes of individuals (males and females) has to be taken into
account in the design of efficient immunization strategies for
sexually transmitted diseases.
\end{abstract}

\maketitle

Disease spreading has been the subject of intense research
since long time ago \cite{MayBook,Cambridge, Murray}.  On the one
hand, epidemiologists have developed mathematical models that can be
used as a guide to understanding how an epidemic spreads and to design
immunization and vaccination policies \cite{MayBook,Cambridge,
Murray}.  On the other hand, data collections have provided
information on the local patterns of relationships in a population. In
particular, persons who may have come into contact with an infectious
individual are identified and diagnosed, making it possible to
contact-trace the way the epidemic spreads, and to validate the
mathematical models.  However, up to a few years ago, some of the
assumptions at the basis of the theoretical models were difficult to
test.  This is the case, for instance, of the complete network of
contacts -the backbone through which the diseases are transmitted.
With the advent of modern society, fast transportation systems have
changed human habits, and some diseases that just a few years ago
would have produced local outbreaks, are nowadays a global threat for
public health systems.  A recent example is given by the severe acute
respiratory syndrome (SARS), that spread very fast from Asia to North
America a few years ago \cite{geisel,guimera,colizza}.  Therefore, it is of utmost importance
to carefully take into account as much details as possible of the
structural properties of the network on which the infection dynamics
occurs.

Strikingly, a large number of statistical properties have been found to be 
common in the topology of real-world social, biological and technological 
networks \cite{albert,siam,PhysRep}.  
Of particular relevance because of its ubiquity
in nature, is the class of complex networks referred to as scale-free (SF)
networks. In SF networks, the number of contacts or connections of a node with
other nodes in the system, the degree (or connectivity) $k$, follows a power
law distribution, $P_k \sim k^{-\gamma}$.  Recent studies have shown the
importance of the SF topology on the dynamics and function of the system under
study \cite{albert,siam,PhysRep}. For instance, SF networks are very robust to
random failures, but at the same time extremely fragile to targeted attacks of
the highly connected nodes \cite{cnsw00,ceah01}.  
In the context of disease spreading, SF contact networks lead to a vanishing 
epidemic threshold in the limit of infinite population when $\gamma \leq 3$
\cite{MaySci,Vespignani,mpv02,newman02}.  This is because the exponent
$\gamma$ is directly related to the first and second moment of the degree
distribution, $\langle k\rangle$ and $\langle k^2\rangle$, and the ratio
$\langle k\rangle / \langle k^2\rangle$ determines the epidemic threshold
above which the outbreak occurs.  When $2 < \gamma \leq 3$, $\langle k\rangle$
is finite while $\langle k^2\rangle$ goes to infinity, that is, the
transmission probability required for the infection to spread goes to zero. 
Conversely, when $\gamma > 3$, there is a finite threshold and 
the epidemic survives only when the spreading rate is above a certain critical  
value. 
The concept of a critical epidemic threshold is central in epidemiology.  
Its absence in SF networks with $2 < \gamma \leq 3$ has a number of important
implications in terms of prevention policies: if diseases can spread and
persist even in the case of vanishingly small transmission probabilities, then
prevention campaigns where individuals are randomly chosen for vaccination are
not much effective \cite{MaySci,Vespignani,mpv02,newman02}.

Our knowledge of the mechanisms involved in disease spreading as well as on
the relation between the network structure and the dynamical patterns of the
spreading process has improved in the last several years
\cite{read1,read2,eubank,eames}. Current approaches are either
individual-based simulations \cite{eubank} or metapopulation models where
network simulations are carried out through a detailed stratification of the
population and infection dynamics \cite{natphys}.  In the particular case of
sexually transmitted diseases (STDs), infections occur within the unique
context of sexual encounters, and the network of contacts
\cite{eames,Liljeros,Lancet,sch,mus,de, freies} is a critical ingredient of
any theoretical framework.  Unfortunately, ascertaining complete sexual
contact networks in significatively large populations is extremely difficult.
However, here we show that it is indeed possible to make use of known global
statistical features to generate more accurate predictions of the critical epidemic
threshold for STDs.

\section{Networks of sexual contacts}
 
Data from national sex surveys \cite{Liljeros,Lancet,sch,mus,de}
provide quantitative information on the number of sexual partners, the
degree $k$, of an individual. Usually, surveys involve a random sample
of the population stratified by age, economical and cultural level,
occupation, marital status, etc.  The respondents are asked to provide
information on sexual attitudes such as the number of sex partners they have had in the last 12 months or 
in their entire life. Although in most cases the
response rate is relatively small, the information gathered is
statistically significant and global features of sexual contact
patterns can be extracted. 
In particular, it turns out that the number of heterosexual partners
reported from different populations is well described by power-law SF
distributions. Table \ref{tab:SCN} summarizes the main results of
surveys conducted in Sweden, United Kingdom, Zimbabwe and Burkina Faso 
\cite{Liljeros,Lancet,sch,mus}.
  
The first thing to notice is the gender-specific difference in the number of
sexual acquaintances \cite{Liljeros,Lancet,sch,mus}. This is manifested by 
the existence of two different exponents in the SF degree distributions, 
one for males ($\gamma_M$) and one for females ($\gamma_F$). 
Interestingly enough, the predominant case in Table \ref{tab:SCN}, 
no matter whether data refers to time frames of 12 months or to entire 
life, consists of one exponent being smaller and the other larger than 3. This is
certainly a borderline case that requires further investigation on the
value of the epidemic threshold.

The differences found in the two exponents $\gamma_F$ and $\gamma_M$ have a
further implication for real data and mathematical modeling.  In an exhaustive
survey, able to reproduce the whole network of sexual contacts, the total
number of female partners reported by men should equal the total number of
male sexual partners reported by women. Mathematically, this means that the
number of links ending at population $M$ (of size $N_M$) equals the number of
links ending at population $F$ (of size $N_F$), which translates into the
following closure relation:
\begin{equation}
N_F\langle
k\rangle_{F}=N_M\langle k\rangle_{M}\;. 
\label{closure}
\end{equation}
Assuming that the degree distributions for the two sets are truly scale-free,
then $P_k^G=\frac{\gamma_G-1}{k_0^{1-\gamma_G}}\cdot k^{-\gamma_G}$, with the
symbol $G$ standing for the gender $(G=F,M)$, and $k_0$ being the minimum
degree. Moreover, if $N_G>>1$ and $\gamma_G>2$ for any $G$,
Eq. (\ref{closure}) gives the relation between the two population sizes as
\begin{eqnarray}
N_M&=&N_F\frac{\langle k\rangle_F}{\langle k\rangle_M}\simeq
N_F\left(\frac{\gamma_M-2}{\gamma_F-2}\right)
\left(\frac{\gamma_F-1}{\gamma_M-1}\right)\;,
\label{scaling}
\end{eqnarray}
which implies that the less heterogeneous (in degree) population must be
larger than the other one.

\begin{table*}
  \caption{{Statistical properties of sexual contact networks from 
  national sex surveys conducted in four different countries: Sweden, United
  Kingdom, Zimbabwe and Burkina Faso. The exponents $\gamma_F$ and $\gamma_M$ are referred
  to the distribution of number of sexual partners cumulated in $12$ months
  and in the respondent's lifetime. The number of respondents is also reported.\label{tab:SCN}}}
\begin{center}
\begin{tabular}{cc|cc|cc|ccc}
  \hline
  \textbf{ survey } & \textbf{ ref. } & \textbf{ $\gamma_F$ (12 months)} &
  \textbf{$\gamma_M$ (12 months)}  & \textbf{$ \gamma_F $ (life)} & \textbf{$ 
  \gamma_M $ (life)} & \textbf{Resp.} & \textbf{Resp. (F)} &
  \textbf{Resp. (M)} \\
  \hline
  \hline
  Sweden & \cite{Liljeros} & $3.54\pm 0.20$ & $3.31\pm 0.20$ & $3.1\pm 0.30$ &
  $2.6\pm 0.30$ & 2810 & - & - \\
  \hline
  U.K. & \cite{Lancet,sch} & $3.10\pm 0.08$ & $2.48\pm 0.05$ & $3.09\pm 0.20$ &
  $2.46\pm 0.10$ & 11161 & 6399 & 4762  \\ 
  \hline
  Zimbabwe & \cite{sch} & $2.51\pm 0.40$ & $3.07\pm 0.20$ & $2.48\pm 0.15$
  & $2.67\pm 0.18$ & 9843 & 5424 & 4419 \\
  \hline
  Burkina Faso & \cite{mus} & $3.9\pm 0.2$ & $2.9\pm 0.1$ & -
  & - & 466 & 179 & 287 \\
  \hline
\end{tabular}
\end{center}
\end{table*}

In conclusion, the empirical observation of two different exponents
demands for a more accurate description of the network of heterosexual
contacts as bipartite SF graphs, i.e. graphs with two set of nodes,
and links connecting nodes from different sets only. In the following
we will consider a graph with $N_M$ nodes, representing males,
characterized by the exponent $\gamma_M$, and $N_M$ nodes,
representing females, characterized by $\gamma_F$.  Concerning the
choice of the couple of exponents from those reported in Table
\ref{tab:SCN}, one must be careful that different STDs have different
associated (recovery) time scales, and that the spreading is based on
the assumption that the links are concurrent on the time scale of the
disease.  In this sense, the exponents extracted from one-year data seem
better suited to most of the STDs, being 
HIV an important exception. However, during the
lifetime of sexually active individuals, sexual behavior is likely to
change due to changes in residence, marital status, age-linked sexual
attitudes, etc \cite{morris}. We thus prefer to use life-cycle data
collections, that integrate all these patterns and can consequently be
regarded as better statistical indicators. After all, the values
reported in Table \ref{tab:SCN} indicate that both one-year and
cumulative data produce exponents in the same range.

\begin{figure}
\begin{center}
\epsfig{file=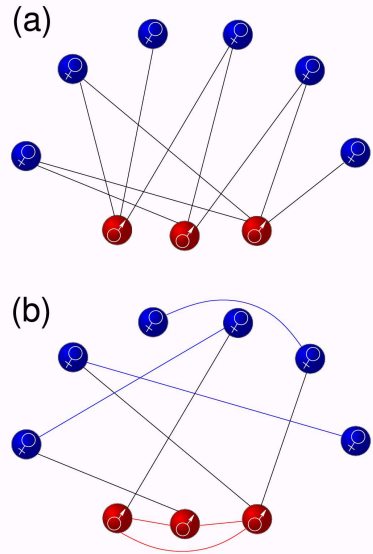, width=6cm, angle=0}
\end{center}
\caption{{\bf Bipartite and unipartite networks.}{\bf (a)} 
Representation of a bipartite network accounting for heterosexual contact 
networks. In such a network we have $N_M$ ($N_F$) nodes representing males
(females), and only male-female links are allowed. Figure {\bf
(b)} represents a rewired version of network {\bf (a)} where the bipartite
nature is lost, while the degree of the nodes is preserved. The two graphs
have the same couple of degree distributions (one for males and one for
females), although only {\bf (a)} reflects the bipartite character of
heterosexual contact networks.}
\label{fig1}
\end{figure}

\section{Theoretical modeling}

The problem of how a disease spreads in a population consisting of two 
classes of individuals can be tackled by invoking the so-called {\em criss-cross}
epidemiological model \cite{Murray}. As illustrated in the bipartite network 
of Fig.~\ref{fig1} (a), in the criss-cross model, the two populations of 
individuals ($N^M$ males and $N^F$
females) interact so that the infection can only pass from one population to
the other by crossed encounters between the individuals of the two
populations, incorporating in this way one of the basic elements of the 
heterosexual spreading of STDs\footnote{We adopted here the indexes $M$ and $F$ to 
denote quantities relative to male and female populations in networks 
of heterosexual contacts. However, the present approach is more general 
and applies to any spreading of diseases in which crossed infections 
between two populations occur.}.
In particular, we consider a {\em susceptible-infected-susceptible} (SIS) 
dynamics, in which individuals can be in one of two different states, namely,
susceptible ($\cal S$) and infectious ($\cal I$).  If  ${\cal S}^M$
and ${\cal I}^M$ (${\cal S}^F$, ${\cal I}^F$) stand for a male (female) respectively in
the susceptible and in the infectious state, the epidemic in the SIS criss-cross
model propagates by the following mechanisms:
\begin{eqnarray}
\label{sis_eq2} 
{\cal S}^F + {\cal I}^M  &\stackrel{\nu_F}{\longrightarrow}& {\cal I}^F + {\cal I}^M\;, 
\nonumber 
\\
{\cal S}^M + {\cal I}^F  &\stackrel{\nu_M}{\longrightarrow}& {\cal I}^M + {\cal I}^F\;,
\label{siseq1} 
\nonumber 
\\
{\cal I}^F   &\stackrel{\mu_F}{\longrightarrow}&   {\cal S}^F, \nonumber 
\\
{\cal I}^{M}   &\stackrel{\mu_{M}}{\longrightarrow}&   {\cal S}^{M}\;, 
\nonumber
\end{eqnarray}
being $\nu_M$, $\nu_F$, $\mu_M$ and $\mu_F$ the infection and recovery
probabilities for males and females. In the case of
heterogeneous contact networks, there is a further compartmentalization of the
population into classes of individuals with the same degree $k$, i.e.  the
same number of sexual partners. Denoting the fraction of males (females) with
degree $k$ in the susceptible or infectious state by $s_k^M$ and $i_k^M$
($s_k^F$ and $i_k^F$), respectively, and adopting a mean field approach
\cite{Murray,Vespignani,mpv02}, the differential equations describing the time
evolution of the densities of susceptible and infected individuals in each
population are
\begin{equation}
\frac{1}{\mu_F}\frac{ d i_{k}^{F}(t)}{d t} =-i_{k}^{F}(t) +\lambda_F k \left[1-i_{k}^{F}(t)
  \right]  \Theta^{M}_{k}(t),
\label{mfk1}
\end{equation}
\begin{equation}
\frac{1}{\mu_M}\frac{ d i^{M}_{k}(t)}{d t} = -i^{M}_{k}(t) +\lambda_M k
\left[1-i^{M}_{k}(t) \right] \Theta^{F}_{k}(t),
\label{mfk2}
\end{equation}
where, $\lambda_G=\nu_G/\mu_G$ ($G=F,M$) are the effective transmission
probabilities.
The quantities $\Theta^{M}_{k}(t)$ , $\Theta^{F}_{k}(t)$ stand for the
probabilities that a susceptible node of degree $k$ 
of one population encounters an 
infectious individual of the other set. Equations (\ref{mfk1}) and
(\ref{mfk2}) have the same functional form of the equation derived in
\cite{Vespignani} for unipartite networks. Neglecting degree-degree
correlations, the critical condition for the occurrence of an endemic state
reduces to:
\begin{equation}
\sqrt{\lambda_F \lambda_M}> \lambda^{*}_{c} = 
\sqrt{\frac{\langle k\rangle_F\langle k\rangle_M}
{\langle k^{2}\rangle_F \langle k^{2}\rangle_M }},
\label{threshold}
\end{equation}
yielding that a necessary condition for the absence 
of the epidemic threshold is the divergence of at least one of
the second moments of the degree distributions, $\langle k^2\rangle_M$ and
$\langle k^2\rangle_F$. Equation (\ref{threshold}) can be compared with the
condition obtained without taking into account that, in heterosexual networks, 
the infection can occur only between male-female couples \cite{Vespignani,MaySci}. 
In fact, working with a unipartite representation of a sexual network, as 
that shown in Fig.~\ref{fig1} (b), with $N_M+N_F$ nodes and a degree
distribution $P_k = ( N_{M} P_k^M + N_{F} P_k^F )/(N_{M}+N_{F})$, one can
express the epidemic threshold as a function of the first and second moments
of the male and female degree distributions as
\begin{equation}
\lambda_c=\frac{\langle k\rangle}{\langle k^2\rangle}
= \frac{2\langle k\rangle_{M}\langle k\rangle_{F}}
{\langle k^2\rangle_{M}\langle k\rangle_{F}+
\langle k^2\rangle_{F}\langle k\rangle_{M}}\;.
\label{threshold2}
\end{equation}
Equations \ (\ref{threshold}) and (\ref{threshold2}) are 
clearly different. For real SF networks of sexual 
contacts, the two thresholds are finite (in the infinity size 
limit) only when the two exponents $\gamma_M$ and $\gamma_F$ are 
both larger than $3$, {\em e.g.} for the one-year number of partners 
in Sweden (see Table \ref{tab:SCN}). 
In such a case, the two expressions read: 
\begin{eqnarray}
\lambda_{c}^{*}&=&\frac{1}{k_0}\sqrt{\frac{(\gamma_F-3)(\gamma_M-3)}{(\gamma_F-2)(\gamma_M-2)}}\;,
\\
\lambda_c&=&\frac{2(\gamma_F-3)(\gamma_M-3)}{k_0\left[(\gamma_F-2)(\gamma_M-3)+(\gamma_M-2)(\gamma_F-3)\right]}\;.
\end{eqnarray}
These two thresholds are only equal in the case $\gamma_M=\gamma_F$
studied in \cite{newman02}.  More importantly, when $\gamma_F \neq
\gamma_M$, we have $\lambda_c^* \ge \lambda_c$, that is the epidemic
state in bipartite networks occurs for larger transmission
probabilities than in unipartite networks.  This result is good news
and highlights the importance of incorporating the crossed infections
scheme in the propagation of STDs.  However, as shown in Table
\ref{tab:SCN}, most of the real networks have at least one exponent
$\gamma_G$ ($G=F,M$) smaller than $3$. This means that, in most of the
practical cases, the two epidemic thresholds vanish as the system size
goes to infinity, no matter the formulation used to model the disease propagation. On the other hand, real populations are finite and thus
the degree distributions have a finite variance regardless of the exponents. Consequently, an epidemic threshold does always exist
and, in order to compare unipartite with bipartite networks, one must
then pay attention to the scaling of the threshold with the size of
the population.

\begin{figure*}
\begin{center}
\epsfig{file=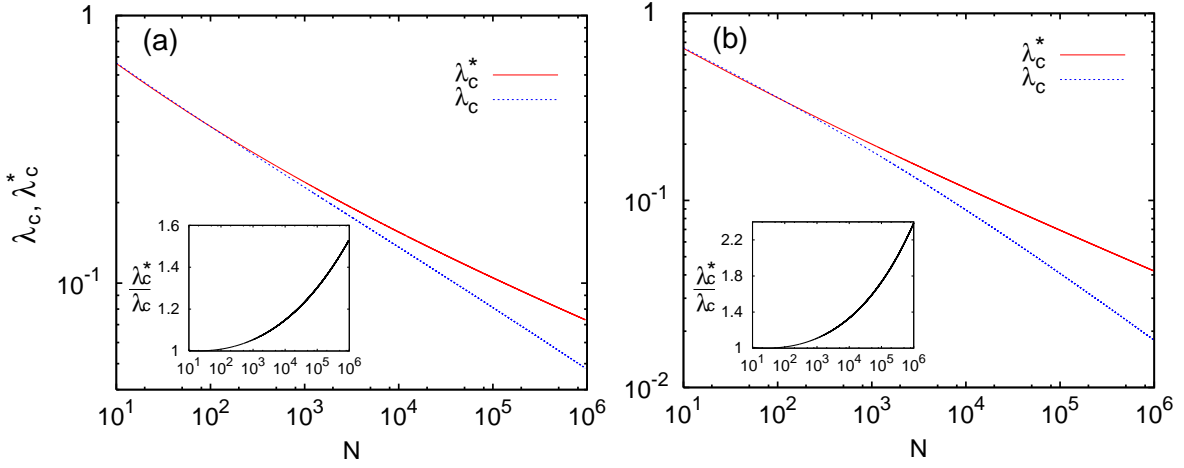, width=16cm, angle=0}
\end{center}
\caption{{\bf Epidemic thresholds as a function of the total population size.}
{The thresholds $\lambda_c^{*}$ (Eq.~(\ref{threshold})) and $\lambda_c$ (Eq. (\ref{threshold2})) obtained for bipartite 
and unipartite networks, are plotted as a function of the population size $N$ for 
two networks with degree distributions as
those found for the sexual networks of {\bf (a)} Sweden \cite{Liljeros} and
{\bf (b)} United Kingdom \cite{Lancet}. The two insets show that
the ratio $\lambda_c^{*}/\lambda_c$ grows as $N$ increases, so that for a
typical population size of $N=10^6$ the thresholds for heterosexual networks
are respectively $53\%$ (Sweden) and $130\%$ (U.K.)  larger than the 
values expected for the unipartite networks.}}
\label{fig2}
\end{figure*}

\subsection{Finite Populations}

We now analyze in more details the differences between $\lambda_c$ and
$\lambda_c^{\star}$ when one of the two exponents (let us say $\gamma_M$ without
loss of generality) is in the range $2<\gamma_M<3$, while
$\gamma_F>2$. 
First, we derive the size scaling of the critical threshold in unipartite
graphs, $\lambda_c$. Equation (\ref{threshold2}) yields
\begin{equation}
\lambda_c=\frac{N_F\langle k\rangle_F+N_M\langle k\rangle_M}{N_F\langle
k^2\rangle_F+N_F\langle k^2\rangle_M}\;.
\nonumber
\end{equation}
Manipulating this expression by considering again
the limit of large (but finite) population sizes, $N_G>>1$ ($G=F,M$), 
we obtain
\begin{equation}
\lambda_c\simeq\frac{2 (3-\gamma_M)/(\gamma_M-2)}
{k_0 \left[N_M^{\frac{3-\gamma_M}{\gamma_M-1}}-1+
\frac{3-\gamma_M}{3-\gamma_F}\frac{\gamma_F-2}{\gamma_M-2}
\left(N_F^{\frac{3-\gamma_F}{\gamma_F-1}}-1\right)\right]}\;.
\nonumber
\end{equation}
The final expression for
$\lambda_c$ can now be obtained by using the closure relation of 
Eq.~(\ref{scaling}) for the $M$ and $F$ population sizes, yielding:
\begin{equation}
\lambda_c\simeq\frac{2(3-\gamma_M)/(\gamma_M-2)}
{k_0\left[N_M^{\frac{3-\gamma_M}{\gamma_M-1}}+
\frac{3-\gamma_M}{3-\gamma_F}
\left(\frac{\gamma_F-2}{\gamma_M-2}\right)^{\frac{2}{\gamma_F-1}}
\left(\frac{\gamma_M-1}{\gamma_F-1}N_M\right)^{\frac{3-\gamma_F}{\gamma_F-1}}
\right]}\;.
\nonumber
\end{equation}
In this formula, only one population size $N_M$ 
appears. Finally, if {\em
e.g.}  $\gamma_F>\gamma_M$, the above equation reduces to
\begin{equation}
\lambda_c\simeq\frac{2(3-\gamma_M)}{k_0(\gamma_M-2)}
N_M^{\frac{\gamma_M-3}{\gamma_M-1}}\;,
\label{l2}
\end{equation}
that contains simultaneously the cases when $2<\gamma_F<3$ and $\gamma_F>3$.

%
%

Now we calculate the scaling of the epidemic threshold in bipartite
(heterosexual) networks, $\lambda_c^*$. 
Manipulating Eq.\ (\ref{threshold}), $\lambda_c^*$ can be
as well expressed as a function of the two exponents 
$\gamma_M$ and $\gamma_F$, and one population size: 
\begin{equation}
\lambda_c^*\simeq\sqrt{\frac{B}{\left(N_M^{\frac{3-\gamma_M}{\gamma_M-1}}-1\right)\left[\left(\frac{\gamma_F-2}{\gamma_M-2}\frac{\gamma_M-1}{\gamma_F-1}N_M\right)^{\frac{3-\gamma_F}{\gamma_F-1}}-1\right]}}\;,
\nonumber
\end{equation}
with $B=\frac{(3-\gamma_M)(3-\gamma_F)}{k_0^2(2-\gamma_M)(2-\gamma_F)}$. The
above expression, when evaluated for $2 < \gamma_G < 3$ ($G=F,M$) and, {\em e.g.},
$\gamma_F>\gamma_M$ yields
\begin{equation}
\lambda_c^*\simeq B^{1/2}\left(\frac{\gamma_F-2}{\gamma_M-2}\frac{\gamma_M-1}{\gamma_F-1}\right)^{\frac{\gamma_F-3}{2(\gamma_F-1)}}N_M^{\frac{1}{2}\left(\frac{\gamma_M-3}{\gamma_M-1}+\frac{\gamma_F-3}{\gamma_F-1}\right)}\;.
\label{eq:lambda*2}
\end{equation}
On the other hand, when {\em e.g.} $\gamma_F>3$, the expression reduces
to
\begin{equation}
\lambda_c^*\simeq\sqrt{\frac{(3-\gamma_M)(\gamma_F-3)}{(2-\gamma_M)(2-\gamma_F)k_0^2}}N_M^{\frac{\gamma_M-3}{2(\gamma_M-1)}}\;.
\label{eq:lambda*}
\end{equation}

\begin{table}
\caption{{\bf Scaling exponents of the epidemic thresholds.}{Scaling
exponents, $\alpha$ and $\alpha^{*}$, of the epidemic thresholds,
$\lambda_c\sim N_{M}^{\alpha}$ and $\lambda_c^{*}\sim N_{M}^{\alpha^{*}}$,
obtained for the SIS model on unipartite networks and when a bipartite network
is considered, respectively. The two situations considered ($2<\gamma_F<3$ and
$\gamma_F>3$) correspond to $2<\gamma_M<3$.\label{tab:EXP}}}
\begin{center}
\begin{tabular}{c|c|c}
\hline
\textbf{ Network } & \textbf{\hspace{15pt}$\alpha^{*}$\hspace{15pt}} & \textbf{$\alpha$}\\
\hline
\hline
$2<\gamma_F<3$ &
$\quad\frac{1}{2}\left(\frac{3-\gamma_F}{\gamma_F-1}+\frac{3-\gamma_M}{\gamma_M-1}\right)\quad$
& $\quad\frac{3-\gamma_M}{\gamma_M-1}\quad$\\
\hline
$\gamma_F>3$   & $\frac{1}{2}\left(\frac{3-\gamma_M}{\gamma_M-1}\right)\quad$ & $\quad\frac{3-\gamma_M}{\gamma_M-1}\quad$\\
\hline
\end{tabular}
\end{center}
\end{table}

\begin{figure*}
\begin{center}
\epsfig{file=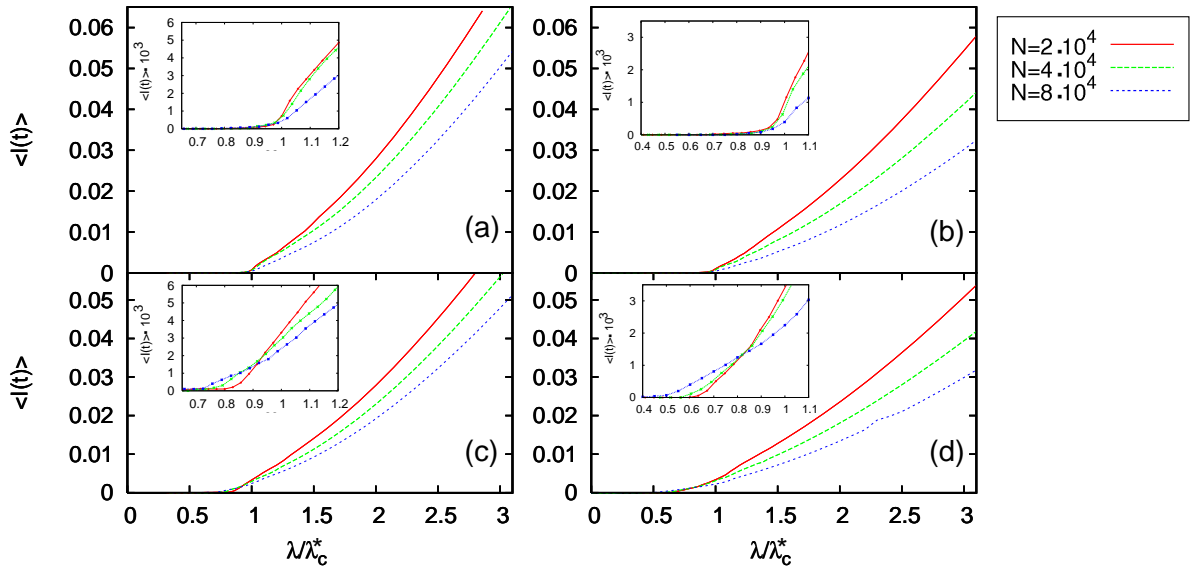, width=16cm, angle=0}
\end{center}
\caption{{\bf Monte Carlo simulations for the UK and Sweden lifetime
networks.}{The SIS phase diagram, $\langle I (t)\rangle$ vs.  $\lambda /
\lambda_{c}^{*}$, is reported for synthetic networks of different sizes. Note
that for each curve the $x$-axis has been rescaled by the theoretical value
(Eq. \ref{threshold}) for the critical point in bipartite networks,
$\lambda_c^*$, of the corresponding network size and exponents $\gamma_F$ and
$\gamma_M$. These exponents used in the network construction (see Methods) are
those extracted from the lifetime degree distributions (see Table
\ref{tab:SCN}) of Sweden \cite{Liljeros}, panels (a) and (c), and
U.K. \cite{Lancet,sch}, panels (b) and (d). The results plotted in (a) and (b)
correspond to the SIS model on a bipartite network, whereas those shown in (c)
and (d) correspond to a unipartite substrate. The numerical results clearly
indicate (see insets) the validity of the analytical predictions for the
epidemic threshold in heterosexual (bipartite) contact networks and the
underestimation of the epidemic threshold when a unipartite substrate is
employed.}}
\label{fig3}
\end{figure*}

\subsection{Comparing the scalings}

Although both epidemic thresholds, $\lambda_c$ and $\lambda_c^*$, tend
to zero as the population goes to infinity, the scaling relations,
$\lambda_c(N_M,\gamma_M,\gamma_F)\sim N_{M}^{-\alpha}$ and
$\lambda_c^{*}(N_M,\gamma_M,\gamma_F)\sim N_{M}^{-\alpha^{*}}$, are
characterized by two different exponents, $\alpha$ and
$\alpha^{*}$. Table \ref{tab:EXP} reports the expression of these two 
exponents as a function of $\gamma_F$ and $\gamma_M$, showing that
$\alpha^{*}$ is always smaller than $\alpha$.  In particular, for the
most common case (see Table \ref{tab:SCN}), i.e.  when one degree
distribution exponent is in the range $]2,3]$, and the other one is
larger than $3$, the value of $\alpha^{*}$ found for bipartite
networks is two times smaller than $\alpha$.  As a consequence, the
results show that in finite bipartite populations the onset of the
epidemic takes place at larger values of the spreading rate. In other
words, it could be the case that for a given transmission probability,
in the unipartite representation shown in Fig.~\ref{fig1} (b) the
epidemic would have survived infecting a fraction of the population,
while when only crossed infections are allowed, as in Fig.~\ref{fig1}
(a), the same disease would not have produced an endemic state.
Moreover, the difference between the epidemic thresholds predicted 
by the two approaches increases with the system size. 
This dependency is shown in 
Fig.~\ref{fig2}, where we have reported, as a function of the system
size, the critical thresholds obtained by numerically solving
Eqs. (\ref{threshold}) and (\ref{threshold2}) with the values of
$\gamma_M$ and $\gamma_F$ found for the lifetime distribution of 
sexual partners in Sweden \cite{Liljeros} and U.K \cite{Lancet,sch}.

\section{Numerical Simulations}

To check the validity of the analytical arguments and also to explore
the dynamics of the disease above the epidemic threshold, we have
conducted extensive numerical simulations of the SIS model in 
bipartite and unipartite computer-generated networks. 
Bipartite and unipartite graphs of a given size are 
built up (see Methods section) having the same degree distributions, 
$P_k^M$ and $P_k^F$, and thus they only 
differ in the way the nodes are linked. A fraction of infected
individuals is initially randomly placed on the network and the SIS
dynamics is evolved: at each time step susceptible individuals get
infected with probability $\nu$ if they are connected to an 
infectious one, and get recovered with  probability $\mu=1$ 
(hence, the effective transmission probability is $\lambda=\nu$). 
After a transient time, the system reaches a
stationary state where the total prevalence of the disease, $\langle
I(t) \rangle$, is measured (see Methods).  The results are finally
averaged over different initial conditions and network realizations.
Fig.~\ref{fig3} shows the fraction of infected individuals as a function of
$\lambda/\lambda_{c}^{*}$ for several system sizes and for the bipartite ((a)
and (b)) and unipartite ((c) and (d)) graphs. In this figure, the infection
probability $\lambda$ has been rescaled by the theoretical value
$\lambda_{c}^{*}$ given by Eq.\ (\ref{threshold}). The purpose of the
rescaling is twofold. First it allows to check the validity of the theoretical
predictions and, at the same time, it provides a clear comparison of the
results obtained for bipartite networks with those obtained for the unipartite
case. Again we have used the values of  $\gamma_M$ and $\gamma_F$ 
extracted from the lifetime number of sexual partners reported for Sweden 
and U.K. \cite{Liljeros,Lancet,sch}. Fig.~\ref{fig3} indicates that 
the analytical solution, Eq.~(\ref{threshold}), is in good agreement with 
the simulation results for the two-gender model formulation. Conversely when the bipartite nature of the underlying graph is not taken into account, the epidemic threshold is 
underestimated, being $\lambda_c/\lambda_c^{*}$ 
smaller than $1$. In addition to this, the error 
in the estimation grows as the population size increases, in agreement with 
our theoretical predictions.

\section{Conclusions}

The inclusion of the bipartite nature of contact networks to describe crossed
infections in the spread of STDs in heterosexual populations is seen to affect
strongly the epidemic outbreak and leads to an increase of the epidemic
threshold. Our results show that, even in the cases when the epidemic
threshold vanishes in the infinite network size limit, the epidemic incidence
in finite populations is less dramatic than actually expected for unipartite
scale-free networks. The results also
point out that the larger the population, the greater the gap between the
epidemic thresholds predicted by the two models, therefore highlighting the
need to accurately take into account all the available information on how
heterosexual contact networks look like. Our results also have important
consequences for the design and refinement of efficient degree-based
immunization strategies aimed at reducing the spread of STDs. In particular,
they pose new questions on how such strategies have to be modified when the
interactions are further compartmentalized by gender and only crossed
infections are allowed.  We finally stress that the present approach is
generalizable to other models for disease spreading ({\em e.g.} the ``{\em
susceptible-infected-removed}'' model) and other processes where
crossed infection in bipartite networks is the mechanism at work.

\section{Methods}

\subsection{Bipartite Network construction}

Synthetic bipartite networks construction starts by fixing the number of
males, $N_{m}$ and the two exponents $\gamma_M$ and $\gamma_F$ of the
power-law degree distributions corresponding to males and females
respectively. The first stage consists of assigning the connectivity $k_i^M$
($i=1$,...,$N_M$) to each member of the male population by generating $N_m$
random numbers with probability distribution $P_k^M =A_M k^{-\gamma_M}$
($\sum_{k_0}^{\infty}A_M k^{-\gamma_M}=1$, with $k_0=3$). The sum of these
$N_m$ random numbers fixes the number of links $N_l$ of the network. The next
step is to construct the female population by means of an iterative
process. For this purpose we progressively add female individuals with a
randomly assigned degree following the distribution $P_k^F =A_F k^{-\gamma_F}$
($\sum_{k_0}^{\infty}A_F k^{-\gamma_F}=1$, with $k_0=3$). Female nodes are
incorporated until the total female connectivity reaches the number of male
edges, $\sum_i k_i^F\leq N_l$. In this way one sets the total number of
females $N_F$. Once the two sets of $N_M$ males and $N_F$ females with their
corresponding connectivities are constructed each one of the $N_l$ male edges
is randomly linked to one of the available female edges avoiding multiple
connections. Finally those few female edges that did not receive a male link
in the last stage are removed and the connectedness of the resulting network
is checked.

\subsection{Unipartite Network construction}

Synthetic unipartite networks has been constructed in two ways. The simplest
one consists of taking the two sets of $N_M$ males and $N_F$ females
constructed for the bipartite network and apply a rewiring process to the
entire population, {\em i.e.} allowing links between individuals of the same
sex. In the second method, a set of $N=N_M+N_F$ individuals whose
connectivities are randomly assigned following the degree distribution
$P(k)=(N_M/N) P_k^M +(N_F/N) P_k^F$ is generated before applying a wiring
process between all pairs of edges. In both methods the wiring process avoids
multiple and self connections and those isolated edges that remain at the end
of the network construction are removed. The connectedness of the networks is
also checked.

\subsection{Numerical Simulations of SIS dynamics}

Montecarlo simulations of SIS dynamics are performed using networks of sizes
ranging from $N=2\cdot 10^4$ to $N=8\cdot 10^4$. The initial fraction of
infected nodes is set to $1\%$ of the network size. The SIS dynamics is
initially evolved for a time typically of $10^4$ time-steps and after this
transient the system is further evolved over consecutive time windows of
$2\cdot 10^3$ steps. In these time windows we monitor the mean value of the number of
infected individuals, $\langle I(t)\rangle$. The steady state is reached if the
absolute difference between the average number of infected individuals of two
consecutive time windows is less than $1/\sqrt{N}$.

\begin{acknowledgments}
We thank K.T.D. Eames and J.M. Read for their useful suggestions. Y.M. is
supported by MEC through the Ram\'{o}n y Cajal Program. This work has been
partially supported by the Spanish DGICYT Projects FIS2006-12781-C02-01 and
FIS2005-00337, and by the Italian TO61 INFN project. 
\end{acknowledgments}

\end{document}